\documentclass[12pt,a4paper]{article}
\usepackage{epsfig}
\def\beq{\begin{equation}}
\def\eeq{\end{equation}}
\def\beqn{\begin{eqnarray}}
\def\eeqn{\end{eqnarray}}
\def\ee{e^+e^-}
\def\yc{y_{\rm cut}}
\def\as{\alpha_{\rm S}}
\def\HW{{\small HERWIG}}
\def\eps{\epsilon}

\begin{document}

\begin{flushright}Cavendish--HEP--10/16\end{flushright}

\begin{center}{\Large\bf QCD Jets and Parton Showers}\footnote{Contribution to
  Proceedings of Gribov-80 Memorial Workshop on Quantum Chromodynamics
  and Beyond, ICTP, Trieste, Italy, 26-28 May, 2010.}
\end{center}
\begin{center}Bryan R. Webber$^*$\\
University of Cambridge, Cavendish Laboratory, \\
J.J.\ Thomson Avenue, Cambridge CB3 0HE, UK\\
$^*$e-mail: brw1@cam.ac.uk
\end{center}

\begin{abstract}
I discuss the calculation of QCD jet rates in $e^+e^-$ annihilation as
a testing ground for parton shower simulations  and jet finding algorithms.
\end{abstract}

\section{Introduction}
\begin{figure}\begin{center}
\epsfig{file=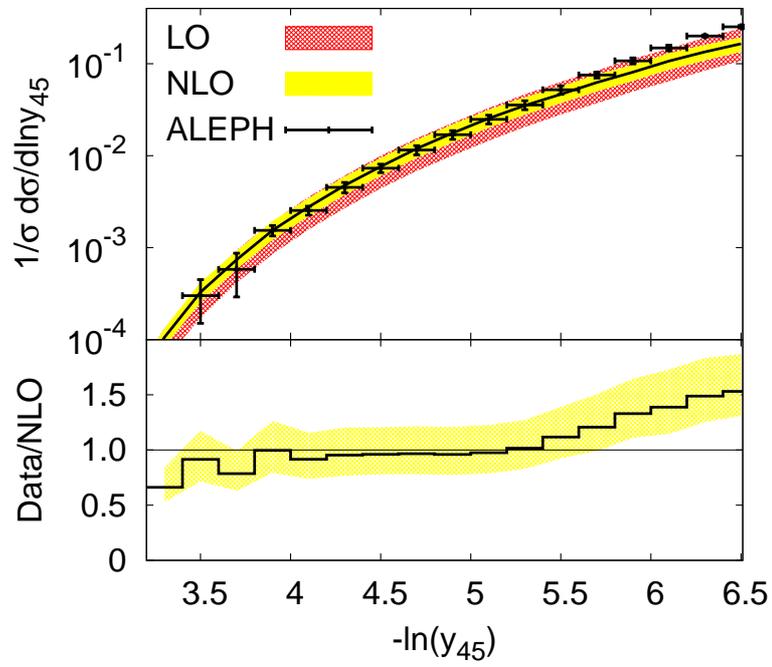,width=0.8\textwidth,angle=270}
\caption{ALEPH data~\cite{Heister:2003aj} on the differential 5-jet
  rate, with the NLO prediction from
  ref.~\cite{Frederix:2010ne}.\label{fig:y45NLO}
} 
\end{center}\end{figure}

The production of jets of hadrons in all kinds of high-energy
collisions is dramatic evidence of the pointlike substructure of
matter. QCD predictions of the rates of production of different
numbers of jets are well confirmed and provide good measurements of
the fundamental coupling $\as$.  The latest triumph in this respect is
the calculation of the 5-jet rate in $e^+e^-$ annihilation to
next-to-leading order, i.e.\ ${\cal O}(\as^4)$~\cite{Frederix:2010ne}.
Figure~\ref{fig:y45NLO} shows that calculation compared to data from the ALEPH experiment
at LEP~\cite{Heister:2003aj}.  The observable shown is $L_{45}\equiv
-\ln(y_{45})$, where $y_{45}$ is the value of the jet resolution
parameter at which five jets are just resolved
using the $k_t$-jet algorithm\cite{Catani:1991hj}.  There is good
agreement over the range shown, and the uncertainty in the prediction
is remarkably small considering this quantity is ${\cal O}(\as^3)$ at leading order.  The
value of the strong coupling obtained from the NLO fit to the region
$L_{45}<6$ is
\beq
\as(M_Z) = 0.1156^{+0.0041}_{-0.0034}\;,
\eeq
which is in good agreement with the world average value obtained from
other observables.

\begin{figure}\begin{center}
\epsfig{file=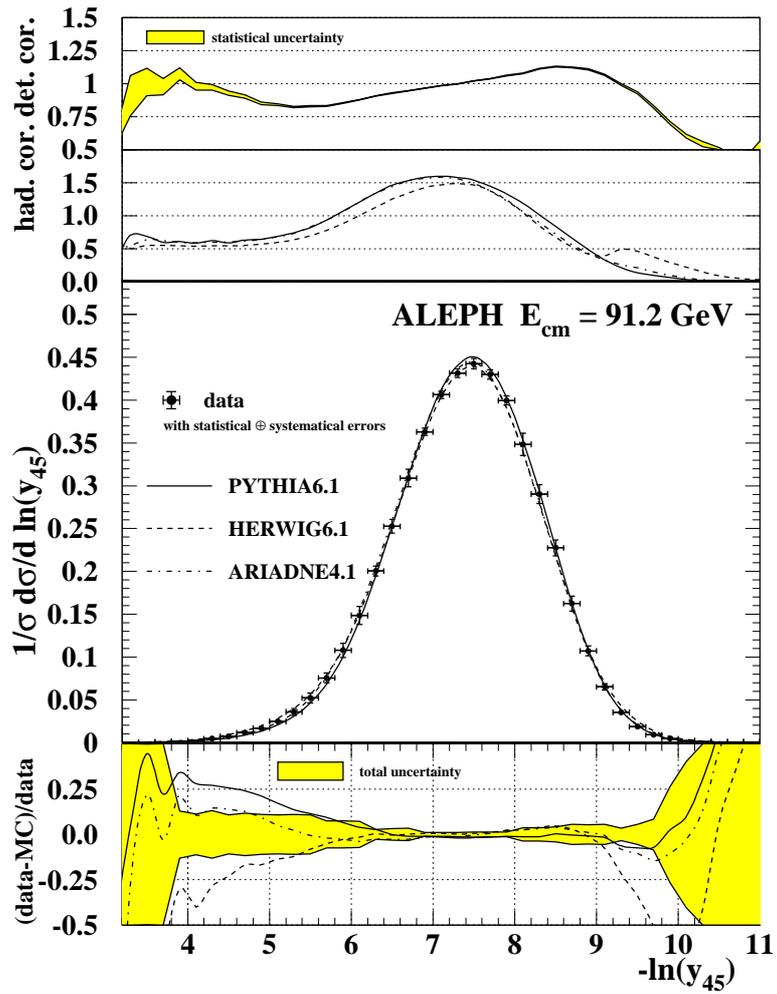,width=0.8\textwidth}
\caption{ALEPH data~\cite{Heister:2003aj} on the differential 5-jet
  rate, with event generator predictions.\label{fig:y45}
}
\end{center}\end{figure}

However, looking at a wider range of $y_{45}$ values,
fig.~\ref{fig:y45}, we see that the region used in the NLO fit
  represents only a small part of the  full distribution.  Most events
  have  $L_{45}>6$, with a distribution that turns over at $L_{45}\sim
  8$, whereas the fixed-order prediction continues to rise more and
  more rapidly with increasing $L_{45}$ (note the logarithmic vertical
  scale in fig.~\ref{fig:y45NLO}).

What this means physically is that most events have a two-jet
structure that can only be resolved into five jets by using a
high-resolution jet algorithm.  However, it is important to understand
this internal structure of the jets as well as possible, for example
to search for highly-boosted new particles whose decays might look
like or be hiding inside QCD jets.  To achieve better understanding we
need progress on two fronts:
\begin{enumerate}
\item Calculations of jet substructure in the region beyond the reach
  of fixed-order perturbation theory;
\item Jet algorithms that probe jets in a way that reveals their
  substructure in informative ways.
\end{enumerate}

Although the era of LEP physics is past, $\ee$ annihilation can still
serve as a good testing ground for ideas on both these topics,
as I hope to illustrate in the following sections.

\section{Parton showers}
The reason for the breakdown of fixed-order predictions at high
$L_{45}$, where most of the data lie, is that QCD matrix elements have
soft and collinear singularities that give rise to logarithmic
enhancement of higher-order contributions.   In fact there are up to
two factors of $L_{45}$ for every extra power of $\as$, so if the
coefficient were unity we would expect a breakdown at $L_{45}\sim
1/\sqrt{\as}\sim 3$.  As we shall see, in fact the coefficient is more
like $2/3\pi$, which does indeed imply a breakdown at $L_{45}\sim 6$.
Ideally we would like to be able to sum these enhanced terms to all
orders in a closed form that would exhibit the turnover in the
distribution, as is the case for several other $\ee$ observables.

\begin{table}
\caption{Jet fractions in $e^+e^-\to$ hadrons to NLL order in
 $L=\ln(1/\yc)$, expanded to third order in $a=\as/\pi$.\label{tab:Rn}}
\begin{tabular}{ccl}
$R_{2 }$ &=& $ 1 + a(R_{21}L + R_{22}L^2) + a^2(R_{23}L^3 + R_{24}L^4) + a^3(R_{25}L^5 + R_{26}L^6) +
\ldots $  \\
$R_{21} $ &=& $ 3C_F/2 $  \\
$R_{22} $ &=& $  -C_F/2 $  \\
$R_{23} $ &=& $ -3C_F^2/4 - 11C_F C_A/36 + C_F N_f /18 $  \\
$R_{24} $ &=& $  C_F^2/8 $  \\
$R_{25} $ &=& $ 3C_F^3/16 + 11C_F^2 C_A/72 - C_F^2 N_f /36 $  \\
$R_{26} $ &=& $ -C_F^3/48 $  \\ \\
$R_{3 }$ &=& $ a(R_{31}L + R_{32}L^2) + a^2(R_{33}L^3 + R_{34}L^4) + a^3(R_{35}L^5 + R_{36}L^6) +
\ldots $  \\
$R_{31} $ &=& $ -3C_F/2  $  \\
$R_{32} $ &=& $ C_F/2 $  \\
$R_{33} $ &=& $ 3C_F^2/2+7C_FC_A/12-C_FN_f/12 $  \\
$R_{34} $ &=& $ -C_F^2/4 - C_F C_A/48 $  \\
$R_{35} $ &=& $ -9C_F^3 /16 - 137C_F^2 C_A/288 - 7C_A^2 C_F /160 + 5C_F^2 N_f /72 + C_F
C_AN_f /160 $  \\
$R_{36} $ &=& $  C_F^3 /16 + C_F^2 C_A/96 + C_F C_A^2 /960 $  \\ \\
$R_{4 }$ &=& $ a^2(R_{43}L^3 + R_{44}L^4) + a^3(R_{45}L^5 + R_{46}L^6) + \ldots $  \\
$R_{43} $ &=& $ -3C_F^2/4-5C_FC_A/18+C_FN_f/36 $  \\
$R_{44} $ &=& $ C_F^2/8 + C_F C_A/48 $  \\
$R_{45} $ &=& $ 9C_F^3 /16 + 71C_F^2 C_A/144 + 217C_F C_A^2 /2880 - 41C_F^2 N_f /720 - C_F
C_AN_f /120 $  \\
$R_{46} $ &=& $ -C_F^3/16-C_F^2C_A/48-7C_FC_A^2/2880  $  \\ \\
$R_{5 }$ &=& $ a^3(R_{55}L^5 + R_{56}L^6) + \ldots $  \\
$R_{55} $ &=& $ -3C_F^3 /16 - 49C_F^2 C_A/288 - 91C_F C_A^2 /2880 + 11C_F^2 N_f /720 + C_F
C_AN_f /480 $  \\
$R_{56} $ &=& $ C_F^3 /48 + C_F^2 C_A/96 + C_F C_A^2 /720 $ 
\end{tabular}
\end{table}

In ref.~\cite{Catani:1991hj} we wrote down integral equations
for generating functions that can be used to compute the leading and
next-to-leading logarithms (NLL) in jet cross sections to any order.
Table~\ref{tab:Rn} shows the results up to ${\cal O}(\as^3)$.  These
equations are for the jet fraction $R_n(\yc)$, which is the fraction
of events that have precisely $n$ jets at resolution $\yc$.  The
differential jet rates, like the one in figs.~\ref{fig:y45NLO} and \ref{fig:y45},
are obtained from them by differentiating:
\beq\label{eq:ykk}
\frac 1{\sigma_{\rm tot}}\frac{d\sigma}{dy_{k-1,k}} =
  -\sum_{n=k}^\infty\left.\frac {dR_n}{d\yc}\right|_{\yc=y_{k-1,k}}\;.
\eeq
Thus to NLL accuracy, in the notation of table~\ref{tab:Rn},
\beq\label{eq:y45NLL}
\frac 1{\sigma_{\rm tot}}\frac{d\sigma}{dy_{45}} =
\frac{a^3}{y_{45}}(6R_{56} L_{45}^5+5R_{55} L_{45}^4) + {\cal O}(\as^4)\;.
\eeq
However, such fixed-order NLL predictions are not much use as they are
invalid when $L_{45}$ is not large and need to be resummed when it is
large.  Indeed, since $6R_{56}=197/270=0.73$ while (for $n_f=5$
flavours) $5R_{55}=-7.77$, the prediction (\ref{eq:y45NLL}) is
actually negative for $L_{45}<10$.

The leading double-logarithmic `abelian' terms, i.e.\ those
proportional to $(a C_F L^2)^{n-2}$, resum to an exponential
form:
\beq\label{eq:RnAb}
R_{n+2}^{(\rm ab)} \sim \frac 1{n!}\left(\frac 12 aC_FL^2\right)^n
\exp \left(-\frac 12 aC_FL^2\right)
\eeq
This gives the correct qualitative features of the
differential distribution (\ref{eq:ykk}) at large $L$, but the
numerical values are wrong, e.g.\  the turn-over occurs at $L_{45}\sim
10$.  This is not surprising in view of the comparable non-abelian
terms and large NLL corrections.

The easiest way to resum the enhanced terms more completely is to
encode them in a parton shower simulation.  By this I mean a
sequential $1\to 2$ parton branching process with branching
probabilities of the form
\beq\label{eq:dP}
dP(a\to bc)= \frac{\as(q')}\pi \frac{dq}q P_{ba}(z)dz
\eeq
where $q$ is an ordered evolution variable, $z$ measures the energy
fraction in the branching, $P_{ba}$ is the corresponding DGLAP
splitting function and the argument $q'$ of $\as$ is a function of $q$
and $z$ in general.  The integral equations of
ref.~\cite{Catani:1991hj} are equivalent to such a process with the
following simple properties: the evolution variable is the angle of
branching and $q'$ is the relative transverse momentum.

The \HW\cite{Corcella:2000bw} event generator results shown in fig.~\ref{fig:y45} are based
on a parton shower with precisely these properties.  {\small PYTHIA}\cite{Sjostrand:2006za}
also has a parton shower which, although organized in a different way,
ought to be equivalent.  {\small ARIADNE}\cite{Lonnblad:1992tz} is
based on a different approach involving colour dipoles rather than
partons.\footnote{I should emphasise that the discussion in this paper
  concerning alternative evolution variables and the colour structure
  of the shower refer only to parton showers as defined by
  eq.~(\ref{eq:dP}) and not to dipole showers.}
All the generators correctly reproduce the main features of
the distribution, in particular the turn-over at $L_{45}\sim 8$.

It should be said that the event generators include a lot of
additional refinements, such as matching to fixed-order matrix
elements at low $L_{45}$ and modelling of hadronization.  In
particular the latter has quite a strong effect at LEP energies
and introduces free parameters which can be tuned to the data.
Nevertheless a parton shower, or equivalent, with the correct features
is an essential component for reliable extrapolation to the higher
energies and different processes encountered at the LHC.

Angular ordering is not the most convenient organization of the parton
shower: physical quantities such as transverse momenta and jet masses
have to be reconstructed from the shower variables.  It would also be
preferable to generate the hardest (highest transverse momentum) branchings
first, which would make matching to fixed-order matrix elements\cite{Catani:2001cc} and
NLO improvements\cite{Frixione:2002ik,Frixione:2007vw} simpler.  These
considerations lead us to look at  what happens if we order the shower
in relative transverse momentum ($p_t$) rather than angle.

Unfortunately with simple $p_t$-ordering things start to go wrong even
at the leading-log level as soon as gluon branching is involved.
Instead of the results in table~\ref{tab:Rn} for the LL coefficients
in the 4-jet and 5-jet fractions,
\beqn 
R_{44}  &=& C_F^2/8 + C_F C_A/48\;,\nonumber\\
R_{56} &=& C_F^3 /48 + C_F^2 C_A/96 + C_F C_A^2 /720\;,
\eeqn 
we get\footnote{Thanks to Mike Seymour for pointing out an error in my
  original calculation of the coefficient of $C_FC_A^2$.}
\beqn\label{eq:Rpt}
R^{(p_t)}_{44} &=& C_F^2/8 + C_FC_A/24\;,\nonumber\\
R^{(p_t)}_{56} &=& C_F^3/48 + C_F^2C_A/48 + 13C_FC_A^2/2880\;.
\eeqn

\begin{figure}\begin{center}
\epsfig{file=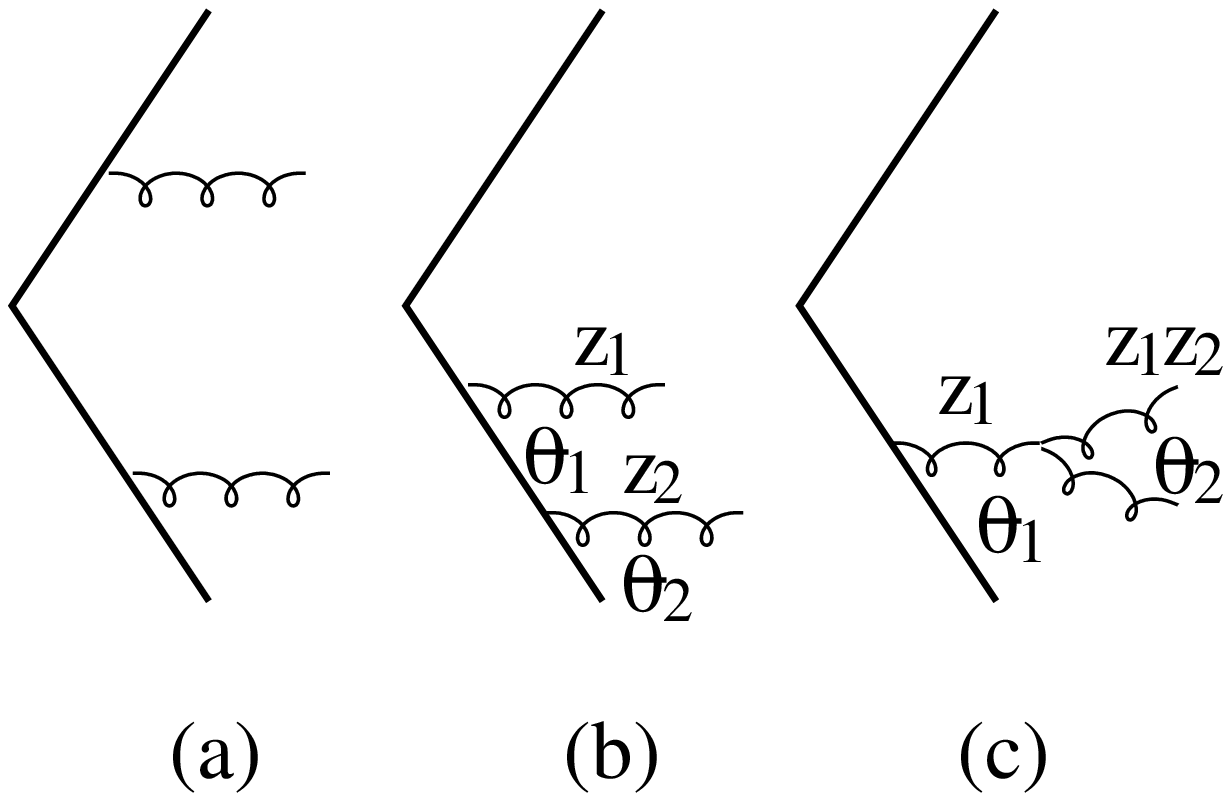,height=4cm}
\caption{Leading order diagrams for $\ee\to$ 4
  jets.\label{fig:fourjets}
}
\end{center}\end{figure}

We could try to fix things up by ordering in $p_t$ and rejecting
branchings that are disordered in angle.  For the 4-jet rate this
cures the problem with gluon branching, fig.~\ref{fig:fourjets}(c),
but spoils the result for sequential quark branching,
fig.~\ref{fig:fourjets}(b), while for the 5-jet fraction everything is wrong:
\beqn
R^{(p_t,\theta)}_{44} &=& 5C_F^2/48 + C_FC_A/48\;,\nonumber\\
R^{(p_t,\theta)}_{56} &=& 7C_F^3/576 + 13C_F^2C_A/1440 + C_FC_A^2/960\;.
\eeqn

\begin{figure}\begin{center}
\epsfig{file=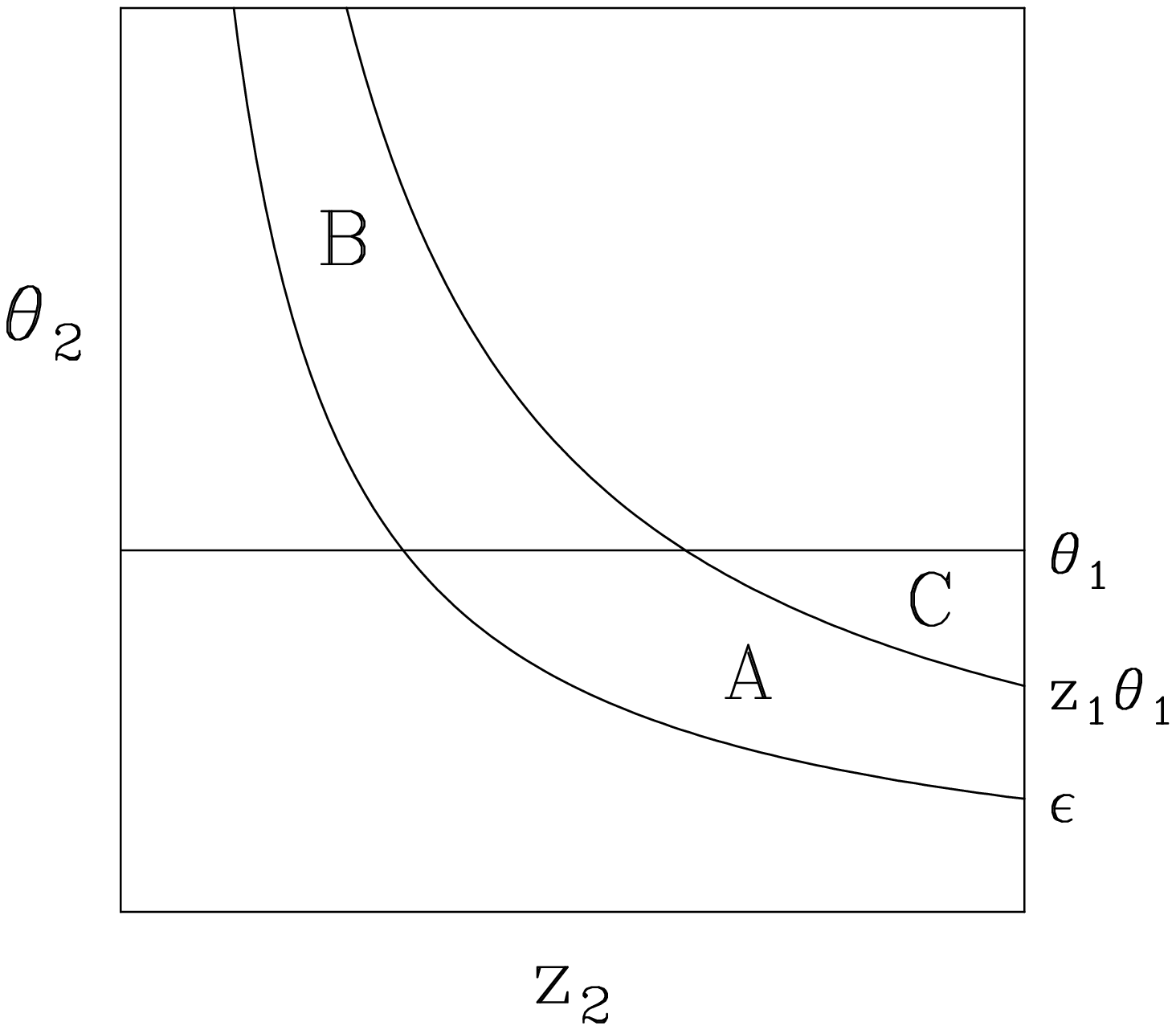,width=0.4\textwidth,angle=90}
\epsfig{file=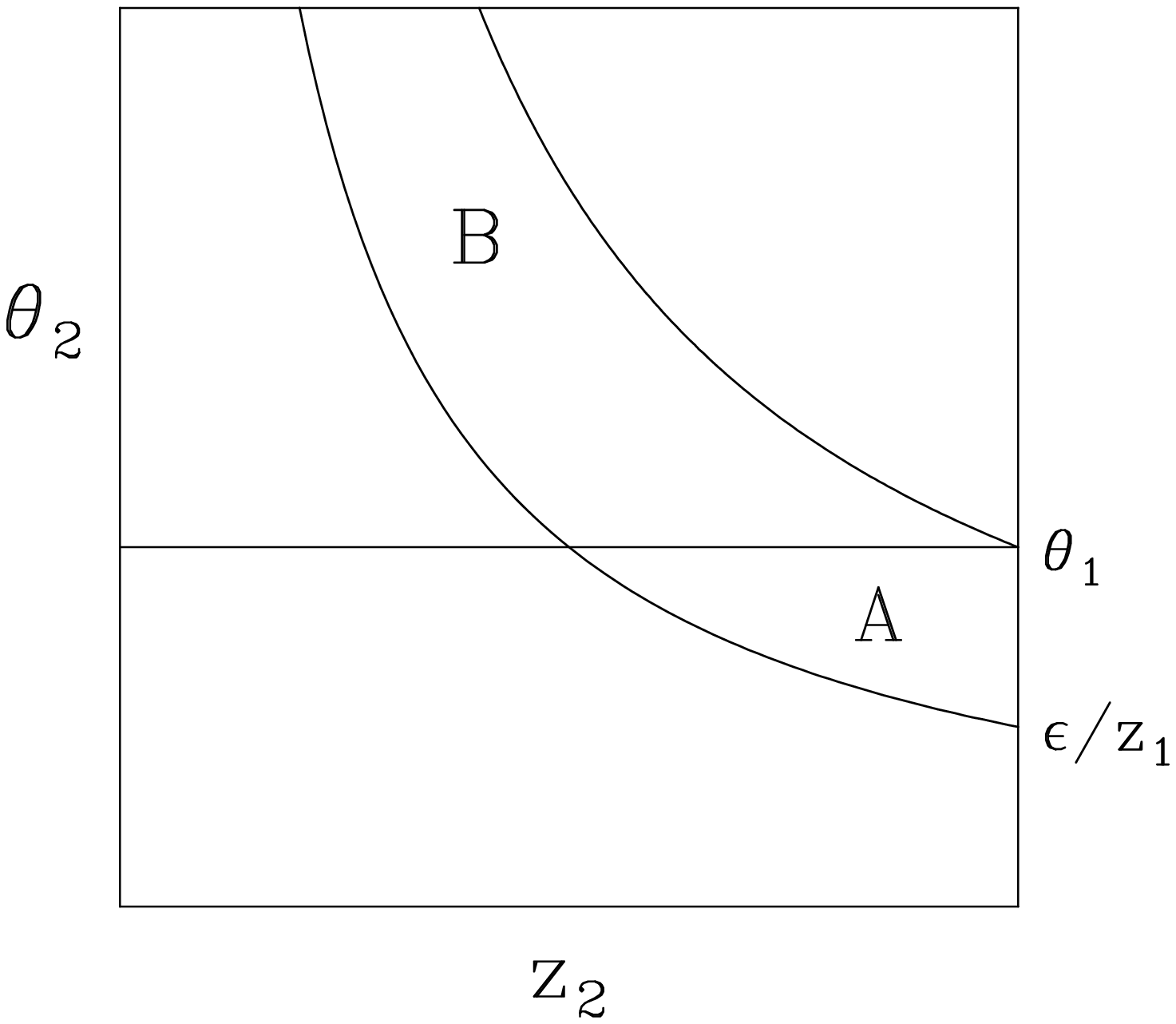,width=0.4\textwidth,angle=90}
\caption{Integration regions for 4-jet diagrams (b) left and (c)
  right.\label{fig:ztheta}
}
\end{center}\end{figure}

To see what is going wrong, consider the $(z_2,\theta_2)$ integration
regions for diagrams \ref{fig:fourjets}(b) and (c), depicted in
fig.~\ref{fig:ztheta}.  Here $z_1,z_2$ and $\theta_1,\theta_2$ are the
(smaller) gluon energy fractions and opening angles in successive
branchings, and $\eps=\sqrt{\yc}$.
Thus in diagram \ref{fig:fourjets}(b), $p_t$-ordering corresponds to
$\eps<z_2\theta_2<z_1\theta_1$, giving the integration region A+B.
However, the correct region is the angular-ordered one A+C.  If we
impose angular ordering after $p_t$-ordering, we get only A, i.e.\ a
deficit in the coefficient of $C_F^2$.

Now it happens that for this diagram the inclusion of region B
compensates for the loss of C as far as the logarithms are
concerned, so in this case $p_t$-ordering alone gives the same result
as angular ordering. I will come back to this point later.

In diagram \ref{fig:fourjets}(c), $p_t$-ordering corresponds to
$\eps<z_1z_2\theta_2<z_1\theta_1$, \i.e. $\eps/z_1<z_2\theta_2<\theta_1$, as
shown on the right in fig.~\ref{fig:ztheta}.   The region C has
disappeared and the $p_t$-ordered region A+B is just too large,
giving an enhanced coefficient of $C_FC_A$.
However, because region C is not there, imposing angular ordering
after $p_t$-ordering is equivalent to simply angular ordering, giving
the correct region A and hence the correct coefficient of $C_FC_A$.

\begin{figure}\begin{center}
\epsfig{file=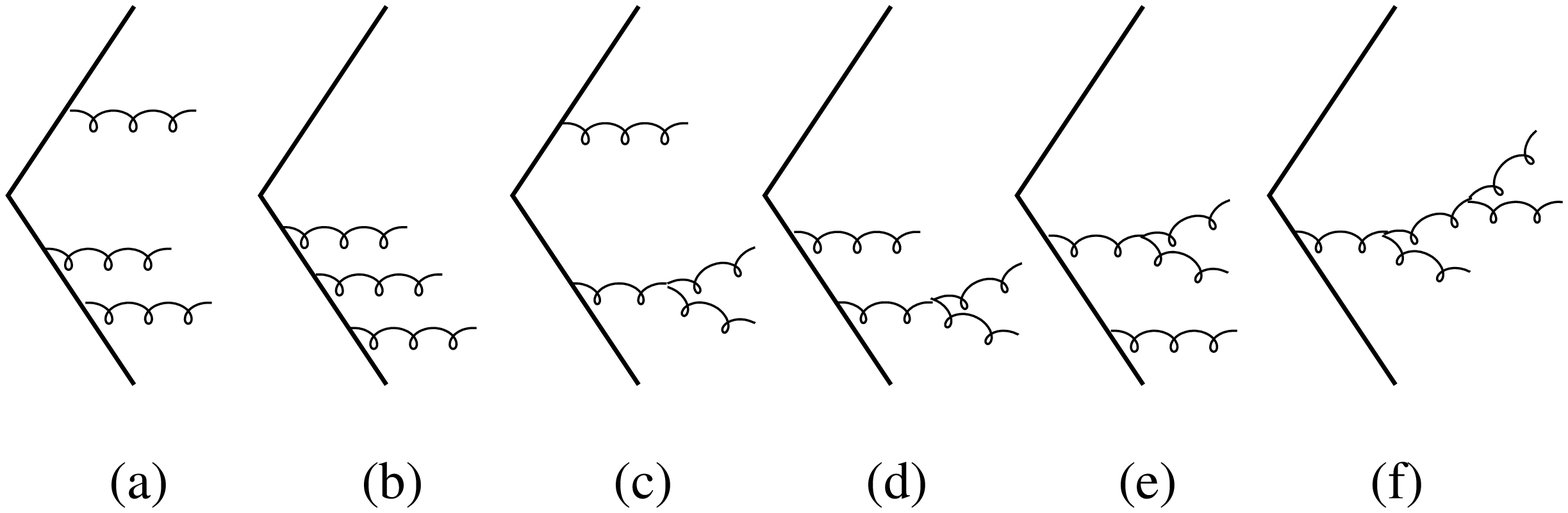,width=0.9\textwidth}
\caption{Leading order diagrams for $\ee\to$ 5
  jets.\label{fig:fivejets}
}
\end{center}\end{figure}

So perhaps the correct prescription for a $p_t$-ordered shower is to
angular-order only the $g\to gg$ vertices?  This corrects the 4-jet
rate but in the 5-jet rate the coefficient of $C_FC_A^2$ is too small:
\beqn
R^{(p_t,gg)}_{44} &=&  C_F^2/8 + C_F C_A/48\;,\nonumber\\
R^{(p_t,gg)}_{56} &=& C_F^3/48 + C_F^2C_A/96 + C_FC_A^2/960\;.
\eeqn
However, the reason for this is the same as before: if the gluon that
branches a second time in fig.~\ref{fig:fivejets}(f)  is the harder
one coming from the first gluon branching, the situation is as on the
left in fig.~\ref{fig:ztheta}, and we should not angular-order the second
gluon branching.

In summary, the way to get the correct LL (and NLL) jet fractions, to
all orders, from a $p_t$-ordered parton shower is to enforce angular ordering with
respect to the branching at which each parton was ``created'', where
this means the branching at which it was the softer of the two
produced~\cite{Catani:2001cc}.  More precisely, one should veto
branchings that are disordered in angle with respect to their
``creation''.  Technically, a veto means not branching but resetting the
$p_t$ scale as if the branching had occurred.  This is a common kind
of procedure in parton shower generators anyway, for example to
correct for flavour thresholds or higher orders in the running coupling.

This looks like a better way to do parton shower event generation.
With $p_t$-ordering one can more easily correct the prediction to NLO,
or indeed to any fixed order in $\as$ in principle.  One only has to
correct the first few steps in the shower.  Unfortunately there is a
catch.  Everything works fine at the parton level as far as the
distribution in phase space is concerned, but the colour
structure of the partonic final state is not correct.

\begin{figure}\begin{center}
\epsfig{file=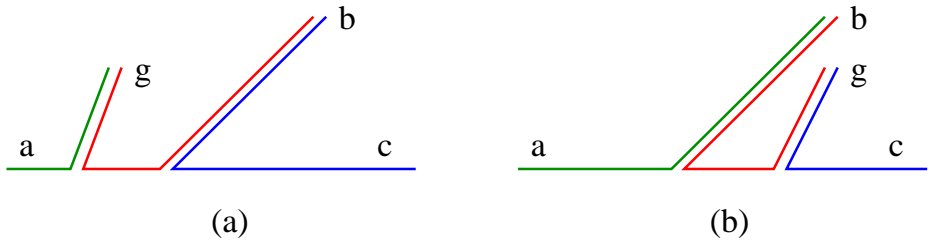,width=0.9\textwidth}
\caption{Large-$N_c$ colour structure of wide-angle gluon emission
  associated with the parton branching $a\to bc$: (a) angular-ordered shower;
(b) $p_t$-ordered shower.\label{fig:colstru}
}
\end{center}\end{figure}

Coming back to fig.~\ref{fig:ztheta} (left), we see that, compared to
angular ordering, $p_t$-ordering includes a region of softer,
wide-angle gluon emission, B, in place of a region of harder, more
collinear emission, C.  What this means is that gluon radiation is
moved around within the shower, the amount and distribution remaining
the same.  This is depicted schematically in fig.~\ref{fig:colstru},
where for simplicity we show the large-$N_c$ approximation, as used
for hadronization in event generators.  In fig.~\ref{fig:colstru}(a), angular ordering
assigns a soft, wide-angle gluon, actually emitted coherently by partons $b$ and $c$, 
to the parent parton $a$, which is reasonable because $a$ does have
the coherent sum of the colour charges of $b$ and $c$.  In contrast,
$p_t$-ordering assigns this gluon to the harder of $b$ and $c$,
in this case $c$, as in fig.~\ref{fig:colstru}(b). That is reasonable as far as the
momenta are concerned, but it spoils the colour structure by treating
$c$ as the colour source and neglecting the coherent contribution of $b$.
 
The colour structure matters when one wants to interface the parton
shower to a non-perturbative hadronization model.  In the cluster
model used by \HW, colour-singlet clusters are formed by
splitting gluons at the end of the shower into $q\bar q$ pairs.
Thus in the angular-ordered fig.~\ref{fig:colstru}(a) the clusters
connect $(gb)$ and $(bc)$, while in $p_t$-ordered
fig.~\ref{fig:colstru}(b) they connect $(bg)$ and $(gc)$.  Similarly
in the {\small PYTHIA} string hadronization model, the string connects $a-g-b-c$ 
in fig.~\ref{fig:colstru}(a) but $a-b-g-c$ in
fig.~\ref{fig:colstru}(b).

In conclusion, an angular-ordered parton shower sums the LL and NLL
enhanced terms and provides partonic final states with colour
structure consistent with QCD coherence.  This is good for
hadronization models but not so convenient for reconstruction of
kinematics or for systematic improvement away from the soft and
collinear regions.  A $p_t$-ordered shower  is better in those
respects and, with the right angular veto procedure, can give the
correct NLL jet fractions. However the colour structure then needs to
be reconfigured according to angular ordering before the partonic
final state can be hadronized.

\section{Jet algorithms}

Recall that the $k_t$-algorithm for $\ee$
annihilation~\cite{Catani:1991hj} is defined in terms of the resolution variable
\beq
y_{ij} = 2\min\{E_i^2,E_j^2\}(1-\cos\theta_{ij})/Q^2\;,
\eeq
where $E_{i,j}$ are the energies of final-state objects $i$ and $j$, $\theta_{ij}$ is
the angle between their momenta and $Q$ is the centre-of-mass energy.
The two objects with the smallest value of $y_{ij}$ are combined into
one, this is repeated until all $y_{ij}>\yc$, and the remaining
objects are called jets.  For the purpose of
counting large logarithms of $\yc$, we can write this in the
small-angle approximation
\beq
\eps_{ij} = \min\{E_i,E_j\}\theta_{ij}/Q >\eps\;,
\eeq
where as before $\eps=\sqrt{\yc}$.

As pointed out in ref.~\cite{Cacciari:2008gp}, this is just one of a
continuum of possible jet algorithms with resolution variable
\beq
\eps_{ij} = \min\{E^p_i,E^p_j\}\theta_{ij}/Q^p\;,
\eeq
where $p$ can be any positive or negative number.  In particular
$p=-1$ defines the resolution for the $\ee$ analogue of the anti-$k_t$
algorithm\cite{Cacciari:2008gp}, which has the advantage that
objects are combined starting with those that have the highest energy rather
than the lowest. 

When $p<0$ a supplementary condition is needed, otherwise infinitely
soft emissions would be resolved.  For anti-$k_t$ we define
\beqn
\eps_{ij} &=&\min\{Q/E_i,Q/E_j\}\theta_{ij}\;,\nonumber\\
\eps_i &=& \eps Q/E_i\;.
\eeqn
Then if the smallest of the set of $\{\eps_{ij},\eps_i\}$ is an
$\eps_i$, we remove $i$ from the list of objects to be recombined, and
if $\eps_i<1$ we call it a jet.  Otherwise we just throw it away.
Thus every jet has an energy greater than $\eps Q$ and is
separated from other jets by an angle greater than $\eps$.
The resulting LL coefficients in the 4- and 5-jet fractions are
\beqn
R^{\rm anti}_{44} &=& C_F^2/2 + C_F C_A/8\;,\nonumber\\
R^{\rm anti}_{56} &=& C_F^3/6 + C_F^2C_A/8 + C_FC_A^2/48\;,
\eeqn
where as before the large logarithm is defined as $L=-2\ln\eps$.
We could introduce an angular resolution $\delta$ different from the
energy resolution $\eps$ by multiplying $\eps_{ij}$ by $\eps/\delta$.
This would just replace $\ln^2\eps$ by $\ln\eps\ln\delta$.

It is easy to see that leading double-logarithmic abelian terms in the
anti-$k_t$ jet rates resum to an exponential form with twice the
exponent of the $k_t$ rates (\ref{eq:RnAb}):
\beq\label{eq:RnAbanti}
R_{n+2}^{(\rm anti, ab)} \sim \frac 1{n!}\left(aC_FL^2\right)^n
\exp \left(-aC_FL^2\right) 
\eeq
It should also be possible to resum the non-abelian and NLL terms using
techniques like those of ref.~\cite{Catani:1991hj}.

\section{Conclusions}
Although the era of high-energy $\ee$ collider experiments is past, at
least for a while, it is helpful to study how our tools for analysing
hadronic final states perform in the cleaner environment of the
annihilation process.

The $k_t$-jet algorithm has proven useful in all kinds of
processes and the $\ee$ jet rates defined in this way are a good place
to test alternative resummation methods, particular those involving
parton showers ordered in different ways.  We have seen that
angular-ordered and $p_t$-ordered showers can both be arranged to
resum the leading and next-to-leading logarithms of the  $k_t$-jet
resolution $\yc$.  The $p_t$-ordering option is good for matching to
fixed-order calculations but causes some difficulties in matching to
hadronization models at low scales, owing to its disordered colour
structure.

The rather different anti-$k_t$ algorithm has been adopted as the
preferred tool for jet finding at the LHC.  An analogous $\ee$
algorithm can be defined and we saw that it has a simple pattern of
leading logarithms, which should be amenable to resummation using
techniques similar to those applied to the $k_t$ algorithm.

\section*{Acknowledgments}
It is a pleasure to recall and acknowledge conversations with
Volodya Gribov in many places during the all-too-brief times we spent
together.  I am also indebted to Stefano Catani, Gavin Salam and Mike
Seymour for helpful comments and discussions.

\end{document}